\documentstyle [preprint,epsf,aps,eqsecnum,floats]{revtex}

\makeatletter

\def\footnotesize{\@setsize\footnotesize{10.0pt}\xpt\@xpt
\abovedisplayskip 10\p@ plus2\p@ minus5\p@
\belowdisplayskip \abovedisplayskip
\abovedisplayshortskip  \z@ plus3\p@
\belowdisplayshortskip  6\p@ plus3\p@ minus3\p@
\def\@listi{\leftmargin\leftmargini
\topsep 6\p@ plus2\p@ minus2\p@\parsep 3\p@ plus2\p@ minus\p@
\itemsep \parsep}}
\long\def\@makefntext#1{\parindent 5pt\hsize\columnwidth\parskip0pt\relax
\def\strut{\vrule width0pt height0pt depth1.75pt\relax}%
$\m@th^{\@thefnmark}$#1}
\long\def\@makecaption#1#2{%
\setbox\@testboxa\hbox{\outertabfalse %
\reset@font\footnotesize\rm#1\penalty10000\hskip.5em plus.2em\ignorespaces#2}%
\setbox\@testboxb\vbox{\hsize\@capwidth
\ifdim\wd\@testboxa<\hsize %
\hbox to\hsize{\hfil\box\@testboxa\hfil}%
\else %
\footnotesize
\parindent \ifpreprintsty 1.5em \else 1em \fi
\unhbox\@testboxa\par
\fi
}%
\box\@testboxb
} %
\global\firstfigfalse
\global\firsttabfalse
\def\tabular{\let\@halignto\@empty\@tabular}
\def\endtabular{\crcr\egroup\egroup $\egroup}
\expandafter \def\csname tabular*\endcsname #1{\def\@halignto{to#1}\@tabular}
\expandafter \let \csname endtabular*\endcsname = \endtabular
\def\@tabular{\leavevmode \hbox \bgroup $\let\@acol\@tabacol
   \let\@classz\@tabclassz
   \let\@classiv\@tabclassiv \let\\\@tabularcr\@tabarray}
\def\endtable{%
\global\tableonfalse\global\outertabfalse
{\let\protect\relax\small\vskip2pt\@tablenotes\par}\xdef\@tablenotes{}%
\egroup
}%

\def\da{d_{\rm A}}

\def\ca{C_{\rm A}}
\def\df{d_{\rm F}}
\def\sf{S_{\rm F}}
\def\stf{S_{2\rm F}}
\def\tr{{\rm tr}}
\def\nf{n_{\rm f}}
\def\MSbar{$\overline{\rm MS}$}
\def\gammaE{{\gamma_{\rm\scriptscriptstyle E}}}
\def\LE{{\cal L}_{\rm E}}
\def\eps{\epsilon}
\def\bball{{\rm ball}}
\def\sun{{\rm sun}}

\def\qcd{{\rm qcd}}
\def\resum{{\rm resum}}
\def\bare{{\rm bare}}

\def\sqed{{\rm sqed}}

\def\PiZ{\Pi^{(0)}}
\def\lnmub{\ln{\bar\mu\over4\pi T}}
\def\alphas{\alpha_{\rm s}}

\def\intd3q{
   \mu^{2\epsilon}\!\!\int {d^{3-2\epsilon}q \over (2 \pi)^{3-2\epsilon}}
}
\def\xbb{{\rm bb}}
\def\xff{{\rm ff}}
\def\xbf{{\rm bf}}
\def\b{{\rm b}}
\def\f{{\rm f}}
\def\Pib{\Pi^\b}
\def\Pif{\Pi^\f}
\def\PifT{\Pi^{\f(T)}}
\def\PibT{\Pi^{\b(T)}}
\def\csch{{\rm csch}}
\def\sgn{{\rm sgn}}

\def\sumint{\hbox{$\sum$}\!\!\!\!\!\!\int}
\def\textsumint{\hbox{\footnotesize $\Sigma$}\!\!\!\hbox{$\int$}}

\begin {document}

\preprint {UW/PT-94-11}

\title{The three-loop free energy for high-temperature QED and QCD
   with fermions}
\author{Peter Arnold}
\address{ Department of Physics, FM-15,
    University of Washington,
    Seattle, Washington 98195
    }%
\author{Chengxing Zhai}
\address{ Department of Physics,
    Purdue University,
    West Lafayette, Indiana 47907
    }%
\maketitle

\date {July 1994}

\begin{abstract}

We compute the free energy density for gauge theories, with fermions,
at high temperature and zero chemical potential.
Specifically, we analytically compute the free energy through $O(g^4)$,
which requires the evaluation of three-loop diagrams.
This computation extends our previous result for pure gauge QCD.

\end{abstract}

\newpage

% ==========================================================================

\section {Introduction}

The perturbative expansion of the free energy of high-temperature gauge theory
has the form
\begin {equation}
  F \sim T^4 [ c_0 + c_2 g^2 + c_3 g^3 + (c'_4 \ln g + c_4) g^4
          + O(g^5) ] \,,
\label {Fform}
\end {equation}
where the $c_i$ are numerical coefficients (with some dependence on the choice
of renormalization scale) and where we have assumed the temperature high
enough that fermion masses can be ignored.
In a previous work \cite{Arnold&Zhai},
we showed how
to compute the coefficient $c_4$ of $g^4$ in pure, non-abelian gauge theory
from three-loop diagrams.  We shall now incorporate fermions into the
theory and so obtain a three-loop result for QED and real QCD.  This
computation is a mostly straightforward extension of our previous work, and so
we refer the reader to that work for motivation and pedagogy.  In fact, the
basic calculations we need to do for fermions very closely parallel those we
did previously for bosons, and our object in this paper will simply be to
point out the relevant differences, catalog results for the basic building
blocks of 3-loop calculations, and present our final results.

In the next section, we fix our notation and conventions for coupling
constants, group factors, and so forth.  In section 2, we outline the
basic integrals that are needed in order to compute fermionic contributions
to the free energy.  In section 3, we show how to derive analytic results
for those basic integrals, though many of the details are left for
appendices.  Finally, in section 4 we present our result for the
free energy and discuss its sensitivity to the choice of renormalization
scale.

% ==========================================================================

\section {Notation and Conventions}

We'll consider gauge theories given by classical Euclidean Lagrangians
of the form
\begin {equation}
   \LE =
   \bar\psi \left( \partial_\mu - i g A^a_\mu T^a \right) \gamma_\mu \psi
   + {1\over4} \left( \partial_\mu A_\nu^a - \partial_\nu A_\mu^a
                + g f^{abc} A_\mu^b A_\nu^c \right)^2
         + (\hbox{gauge fixing}) \,,
\end {equation}
where the $T^a$ are the generators of a single, simple Lie group, such
as U(1) or SU(3).
To simplify presentation, we will not derive results for an arbitrary
product of simple Lie groups such as SU(2)$\times$U(1), but such cases
could easily be handled simply by adjusting the overall group and coupling
factors on the results we give for individual diagrams.
$\da$ and $\ca$ are the dimension and quadratic Casimir of the adjoint
representation, with $\ca$ given by
\begin {equation}
   f^{abc}f^{dbc} = \ca \delta^{ad} \,.
\end {equation}
$\df$ is the dimension of the total fermion representation ({\it e.g.}\ 18 for
six-flavor QCD), and $\sf$ and $\stf$ are defined in terms of the
generators $T^a$ for the total fermion representation as
\begin {equation}
   \sf = {1\over\da} \tr (T^2) \,,
   \qquad
   \stf = {1\over\da} \tr \bigr[ (T^2)^2 \bigl] \,,
\end {equation}
where $T^2 = T^a T^a$.
For SU(N) with $\nf$ fermions in the fundamental representation, the
standard normalization of the coupling gives
\begin {equation}
   \da = N^2-1 \,,
   \qquad
   \ca = N \,,
   \qquad
   \df = N \nf \,,
   \qquad
   \sf = {1\over2} \nf \,,
   \qquad
   \stf = {N^2{-}1 \over 4N} \nf \,.
\end {equation}
For U(1) theory, relabel $g$ as $e$ and let the charges of the $\nf$ fermions
be $q_i e$.  Then
\begin {equation}
   \da = 1 \,,
   \qquad
   \ca = 0 \,,
   \qquad
   \df = \nf \,,
   \qquad
   \sf = \sum_i q_i^2 \,,
   \qquad
   \stf = \sum_i q_i^4 \,.
\end {equation}

We shall work in Feynman gauge.
We also work exclusively in the Euclidean (imaginary time) formulation of
thermal field theory.  We shall conventionally refer to four-momenta
with capital letters $K$ and to their components with lower-case letters:
$K=(k_0,\vec k)$.  All four-momenta are Euclidean with discrete frequencies
$k_0 = 2\pi n T$ for bosons and ghosts and
$k_0 = 2\pi \left(n{+}{1\over2}\right)T$ for
fermions.  We regularize the theory by working in $d=4{-}2\eps$ dimensions
with the modified minimal subtraction (\MSbar) scheme, which corresponds to
doing minimal subtraction (MS) and then changing the MS scale $\mu$ to the
\MSbar\ scale $\bar\mu$ by the substitution
\begin {equation}
   \mu^2 = {e^{\gammaE} \bar\mu^2 \over 4\pi} \,.
\end {equation}
To denote summation over discrete loop frequencies and integration over
loop three-momenta, we use the short-hand notation
\begin {equation}
   \sumint_P \to
   \mu^{2\eps} T \sum_{p_0} \int {d^{3-2\eps} p \over (2\pi)^{3-2\eps}}
\end {equation}
for bosonic momenta and
\begin {equation}
   \sumint_{\{P\}} \to
   \mu^{2\eps} T \sum_{\{p_0\}} \int {d^{3-2\eps} p \over (2\pi)^{3-2\eps}}
\end {equation}
for fermionic momenta, where
\begin {equation}
   \sum_{p_0} \to \sum_{p_0 = 2\pi n T} \,,
   \qquad\qquad
   \sum_{\{p_0\}} \to \sum_{p_0 = 2\pi\left(n{+}{1\over2}\right)T} \,.
\end {equation}
We shall also sometimes use the notation
\begin {equation}
   \sumint_{P+\{P\}} \to \sumint_P + \sumint_{\{P\}} \,.
\end {equation}

\begin {figure}
\vbox
    {%
    \begin {center}
	\leavevmode
	
 	\epsfbox [100 220 500  550] {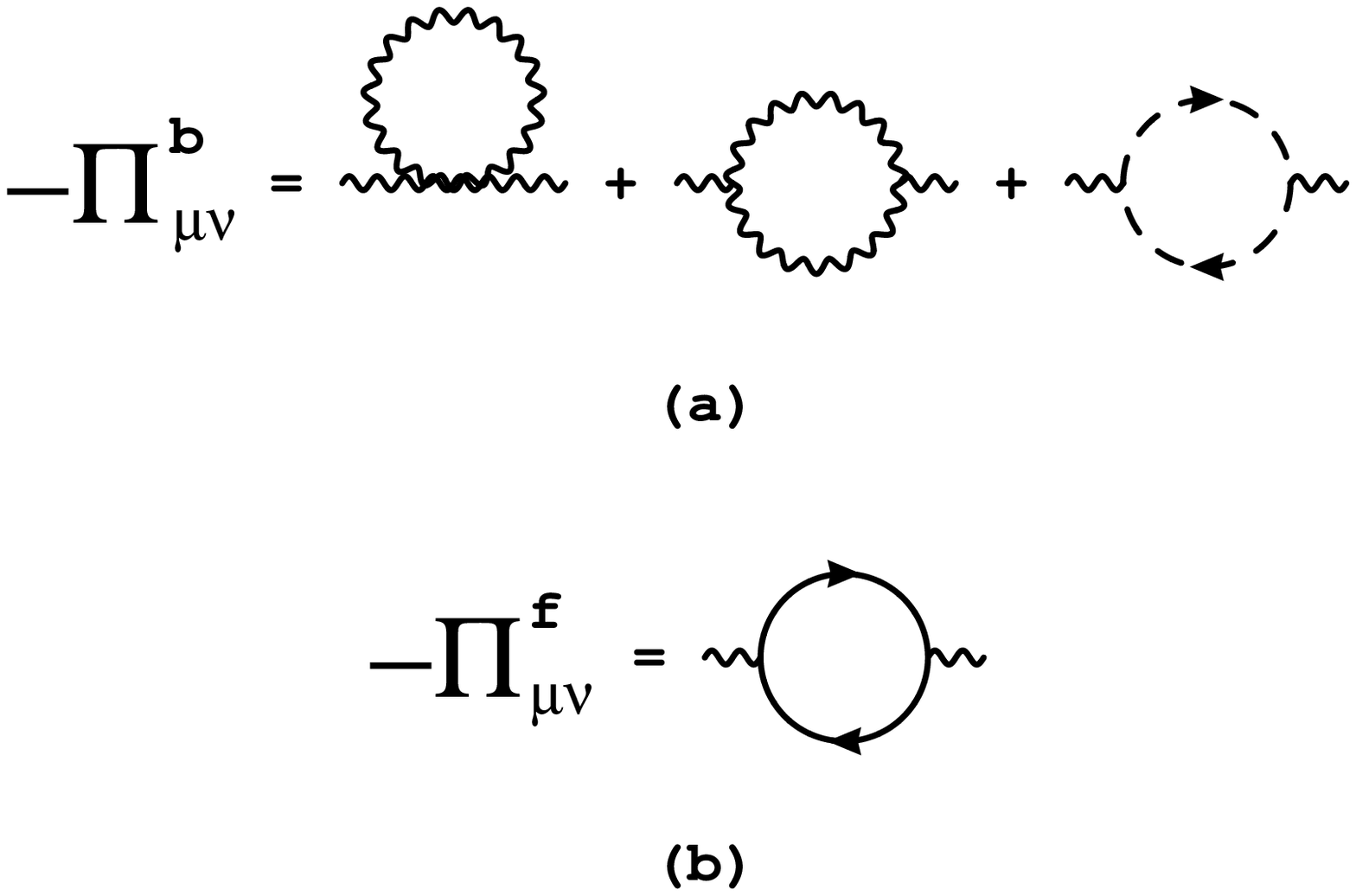}
    \end {center}
    \caption
	{%
	\label {fig pi}
        The (a) bosonic and (b) fermionic contributions to the one-loop
        gluon self-energy.
	}%
    }%
\end {figure}

We handle the resummation of hard thermal loops (which is required to
make perturbation theory well-behaved beyond $O(g^2)$) as we did
in ref.~\cite{Arnold&Zhai}.
Specifically, we must improve our propagators by incorporating the
Debye screening mass $M$ for $A_0$, which is determined at leading order
by the self-energy diagrams of fig.~\ref{fig pi}:
\begin {equation}
   M^2 \delta^{ab}
       = \Pi^{ab}_{00}(0)
       = \Pi^{ab}_{\mu\mu}(0)
       = g^2 \delta^{ab} \left[
               \ca (d-2)^2 \sumint_Q {1\over Q^2}
               - 4 \sf (d-2) \sumint_{\{Q\}} {1\over Q^2}
         \right] \,.
\label{M def}
\end {equation}
This is accomplished
by rewriting our Lagrangian density, in frequency space, as
\begin {equation}
  \LE = \left(\LE + {\textstyle{1\over2}} M^2 A^a_0 A^a_0
                               \delta_{p_0^{}}\right)
                  - {\textstyle{1\over2}} M^2 A^a_0 A^a_0
                               \delta_{p_0^{}} \,,
\label {gauge resummation}
\end {equation}
where $\delta_{p_0^{}}$ is shorthand for the the Kronecker delta function
$\delta_{p_0^{},0}$.
Then we absorb the first $A_0^2$ term into our
unperturbed Lagrangian ${\cal L}_0$
and treat the second $A_0^2$ term as a perturbation.

% ==========================================================================

\section {The Basic Integrals}

The most basic one-loop integrals that appear in high-temperature field
theory are of the form
\begin {equation}
   b_n \equiv \sumint_P {1\over P^{2n}} \,,
   \qquad
   f_n \equiv \sumint_{\{P\}} {1\over P^{2n}} \,.
\end {equation}
The bosonic form of these integrals needed for the calculation were reviewed
in ref.~\cite{Arnold&Zhai} and are given by
\begin {eqnarray}
   b_1 &=&
       {T^2\over12} \left[ 1 + \eps \left(
          2\lnmub + 2 {\zeta'(-1) \over \zeta(-1)} + 2
        \right) \right] + O(\eps^2) \,,
\label {b1 eqn}
\\
   b_2 &=&
       {1\over(4\pi)^2} \left[
	{1 \over \epsilon} + 2\lnmub + 2 \gammaE \right ]
	+ O(\eps) \,.
\label {b2 eqn}
\end {eqnarray}
As noted in ref.~\cite{Coriano&Parwani},
the $f_n$ are then easily determined by
considering $f_n+b_n$ and then scaling the momenta $(p_0,\vec p)$ by
2 so that it becomes proportional to $b_n$.  One finds
\begin {equation}
   f_n = (2^{2n+1-d}-1) b_n \,.
\label {f eqn}
\end {equation}

\begin {figure}
\vbox
    {%
    \begin {center}
	\leavevmode
	
	\epsfbox [100 250 500 470] {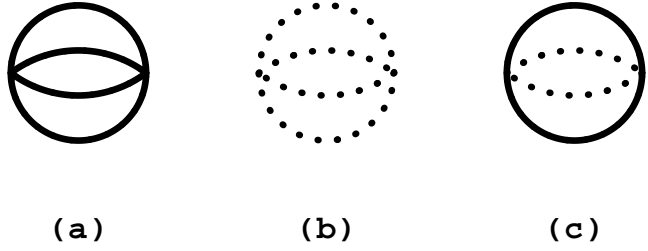}
    \end {center}
    \caption
	{%
	\label {fig bball}
        The (a) bosonic, (b) fermionic, and (c) mixed scalar basketball
        diagrams.  Solid (dotted) lines correspond to scalar propagators
        with bosonic (fermionic) frequencies.
	}%
    }%
\end {figure}

In ref.~\cite{Arnold&Zhai},
we reviewed how three-loop diagrams contributing to the
free energy can be reduced to some simple sum-integrals at $O(g^4)$.
The most basic was
\begin {equation}
   I_\bball^\xbb \equiv \sumint_{PQR} {1\over P^2 Q^2 R^2 (P+Q+R)^2}
   = \sumint_P \left[\Pib(P)\right]^2 \,,
\end {equation}
which corresponds to the basketball diagram of scalar theory, depicted
in fig.~\ref{fig bball}(a).  $\Pib$ is defined by
\begin {equation}
   \Pib(P) \equiv \sumint_Q {1 \over Q^2 (P+Q)^2} \,,
\end {equation}
and we have introduced the superscript b for $\Pi$ to indicate that it
is defined with a bosonic frequency sum.  When fermions are included,
the reduction of diagrams to a few simple integrals requires introducing
some fermionic relatives of $I_\bball^\xbb$:
\begin {eqnarray}
   I_\bball^\xff &\equiv& \sumint_P \left[\Pif(P)\right]^2 \,,
\\
   I_\bball^\xbf &\equiv& \sumint_P \Pib(P) \Pif(P) \,,
\end {eqnarray}
\begin {equation}
   \Pif(P) \equiv \sumint_{\{Q\}} {1\over Q^2 (P+Q)^2} \,.
\end {equation}
These are depicted by figs.~\ref{fig bball}(b) and (c).

\begin {figure}
\vbox
    {%
    \begin {center}
	\leavevmode
	
 	\epsfbox [100 270 500 500] {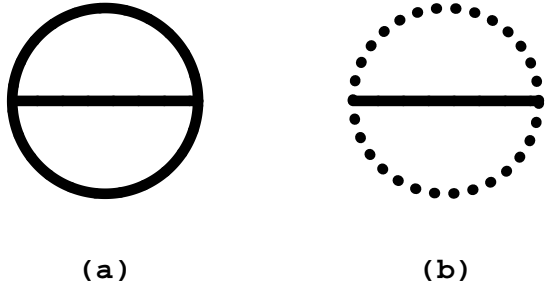}
    \end {center}
    \caption
	{%
	\label {fig sun}
        The (a) bosonic and (b) fermionic setting sun diagrams.
	}%
    }%
\end {figure}

Another basic integral encountered in the pure gauge theory case was
the one associated with the scalar sunset diagram of fig.~\ref{fig sun}(a),
evaluated to leading order in masses:
\begin {equation}
   I_\sun^\b (m_1,m_2,m_3) \equiv
   \sumint_{PQ} {1\over (P^2+m_1^2) (Q^2+m_2^2) [(P+Q)^2+m_3^2]} \,.
\end {equation}
When fermions are included, one also needs
\begin {equation}
   I_\sun^\f \equiv
   \sumint_{\{PQ\}} {1\over P^2 Q^2 (P+Q)^2}
   = \sumint_P {1\over P^2} \Pif(P) \,,
\end {equation}
corresponding to fig.~\ref{fig sun}(b).
The above integral is infrared finite because fermionic Euclidean frequencies
$p_0$ are never zero; so, unlike the bosonic case, the masses can be dropped
at leading order in $m/T$.  By using the same contour-trick argument
that was used in Appendix F.1 of ref.~\cite{Arnold&Zhai},
one can easily show that
\begin {equation}
   I_\sun^\f = 0 \,.
\end {equation}
However, it is sometimes convenient to also know the pieces of $I_\sun^\f$
corresponding to restricting the frequency sum in various ways, and these are
discussed in Appendix~\ref{sunset appendix}.

\begin {figure}
\vbox
    {%
    \begin {center}
	\leavevmode
	
	\epsfbox [100 100 500 700] {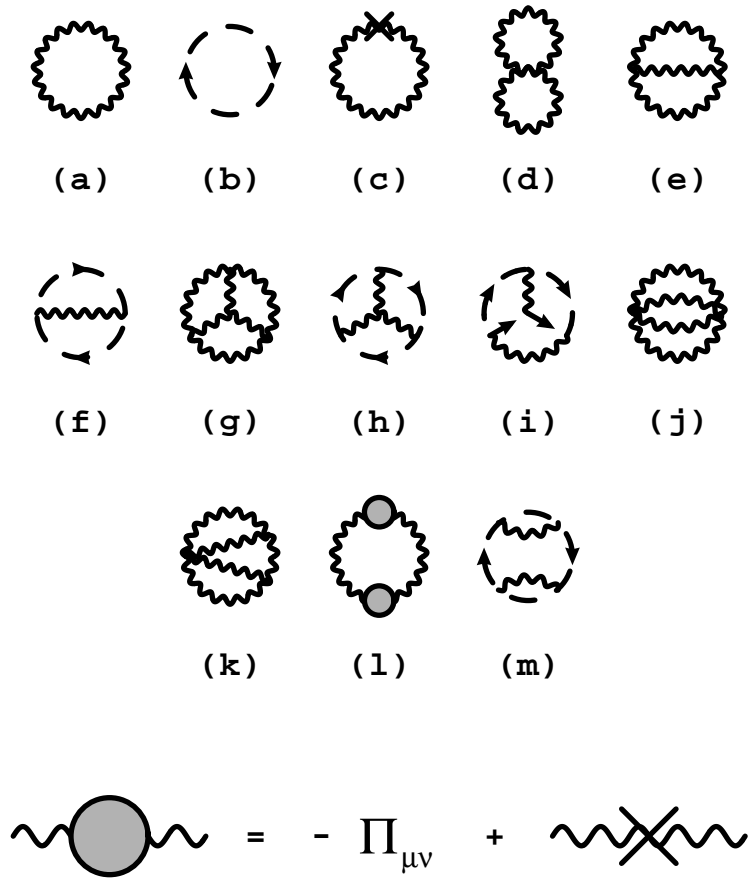}
    \end {center}
    \caption
	{%
	\label {figf}
        Diagrams contributing to the free-energy in pure gauge theory.
        When fermions are added, we include the fermionic contribution
        to $\Pi_{\mu\nu}$ in diagram (l) and include the diagrams of
        fig.~\protect\ref{figf2}.
        The crosses are the ``thermal counter-terms'' arising from the
        last term of (\protect\ref{gauge resummation}),
        and the dashed lines are ghosts.
        We have not explicitly shown any zero-temperature counter-terms, and
        each diagram should be multiplied by the appropriate
        multiplicative renormalizations for vertices and propagators.
	}%
    }%
\end {figure}

Finally, the pure gauge theory calculation required the integral
corresponding to the bosonic piece of fig.~\ref{figf}(l):
\begin {equation}
   \da \ca^2 g^4 I_\qcd^\xbb \equiv
   \sumint_P {1\over P^4} \tr\,[ \Delta\Pi^\b_{\mu\nu}(P) ]^2 \,,
\end {equation}
where $\Pi^\b_{\mu\nu}$ is the bosonic contribution to the vector self-energy,
given by fig.~\ref{fig pi}(a), and the notation
\begin {equation}
   \Delta\Pi_{\mu\nu}(P) \equiv \Pi_{\mu\nu}(P)
   - \Pi_{\mu\nu}(0) \delta_{p_0^{}} \,.
\end {equation}
has been used.

We shall need the same integral with the complete
self-energy, which means we need
\begin {eqnarray}
   \da \sf^2 g^4 I_\qcd^\xff &\equiv&
   \sumint_P {1\over P^4} \tr\,[ \Delta\Pi^\f_{\mu\nu}(P) ]^2 \,,
\\
   \da \ca \sf g^4 I_\qcd^\xbf &\equiv&
   \sumint_P {1\over P^4} \tr\, \Delta\Pi^\b_{\mu\nu}(P)
                              \Delta\Pi^\f_{\mu\nu}(P)  \,,
\end {eqnarray}
where $\Pi_{\mu\nu}^\f$ is the fermionic
contribution given by fig.~\ref{fig pi}(b).

In the pure gauge theory case, all three-loop diagrams except
fig.~\ref{figf}(l)
could be reduced to the basketball integral $I_\bball^\xbb$
by the application of a few simple tricks.  For instance,
fig.~\ref{figf}(i) is equal to
\begin{equation}
    - {1 \over 8} \da \ca^2 g^4 \sumint_{PQK}
	{P \cdot (Q-K) \, (P-K) \cdot Q
	\over P^2 Q^2 K^2 (P-Q)^2 (Q-K)^2 (K-P)^2}
\label {reduction example}
\end{equation}
and is reduced by (1) expanding numerator factors in terms of
denominator factors to cancel factors between numerator and denominator,
such as
\begin {equation}
    P\cdot(Q-K) = {\textstyle{1\over2}} [(K-P)^2 - K^2 - (P-Q)^2 + Q^2] \,;
\label{(i) expansion}
\end {equation}
(2) performing appropriate changes of variables to collect similar terms;
and (3) using the identity
\begin {equation}
   \sumint_P {P_{\mu} \over (P+Q)^2 (P+K)^2}
	= - {Q_{\mu} + K_{\mu} \over 2}
	\sumint_P {1 \over (P+Q)^2 (P+K)^2} \,.
\label {trick}
\end {equation}
The last identity follows by changing variables $P{\to}{-}P{-}Q{-}K$,
\begin {equation}
   \sumint_P {P_{\mu} \over (P+Q)^2 (P+K)^2}
	= -  \sumint_P {P_{\mu} \over (P+Q)^2 (P+K)^2}
          - {(Q_{\mu} + K_{\mu})}
	           \sumint_P {1 \over (P+Q)^2 (P+K)^2} \,,
\end {equation}
and then moving the first-term on the right-hand side over to the
left-hand side.  Unfortunately, this trick does not generalize to the
case where $Q{+}K$ is fermionic instead of bosonic.  If $Q{+}K$ is
fermionic, then
\begin {equation}
   \sumint_P {P_{\mu} \over (P+Q)^2 (P+K)^2}
	= -  \sumint_{\{P\}} {P_{\mu} \over (P+Q)^2 (P+K)^2}
          - {(Q_{\mu} + K_{\mu})}
	           \sumint_{\{P\}} {1 \over (P+Q)^2 (P+K)^2} \,,
\end {equation}
and there is no simple way to solve for the bosonic integral on the
left-hand side.  The failure of this trick requires us to introduce a
new fundamental integral, as was done by Parwani and Corian\`o
in ref.~\cite{Coriano&Parwani}:%
\footnote{
   We have adopted their notation, $H_3$, for this integral.
   Their $H_1$ and $H_2$ correspond to our $I_\bball^\xbb$ and
   $I_\bball^\xff$ respectively, and their $H_4$ is discussed in
   Appendix~\ref{qcd appendix}.
}
\begin {equation}
   H_3 \equiv \sumint_{\{P\}QK}
    {Q \cdot K \over P^2 Q^2 K^2 (P+Q)^2 (P+K)^2} \,.
\label {H3 def}
\end {equation}
If $P$ were bosonic, this would be reducible by (\ref{trick}).

\begin {figure}
\vbox
    {%
    \begin {center}
	\leavevmode
	
 	\epsfbox [100 300 500 450] {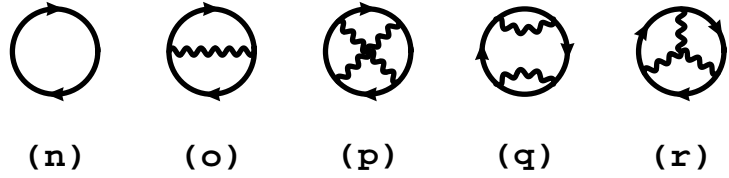}
    \end {center}
    \caption
	{%
	\label {figf2}
        Diagrams that must be added to fig.~\protect\ref{figf} to include
        fermions in the calculation of the free energy.
	}%
    }%
\end {figure}

Figs.~\ref{figf} and \ref{figf2}
show all of the diagrams contributing to the free-energy up
to three loops.  The reductions of all the three-loop diagrams to the basic
integrals are given in Appendix~\ref{graphs appendix}.

% ==========================================================================

\section {Results for Integrals}

\subsection{The fermionic basketball $I_\bball^\xff$}

   The derivation of $I_\bball^\xff$ closely parallels that of the
bosonic $I_\bball^\xbb$ in ref.~\cite{Arnold&Zhai} with the main difference
being that the bosonic sum identity
\begin {equation}
   \sum_{q_0} e^{-|q_0|r} e^{-|p_0+q_0|r}
      = (\coth \bar r + |\bar p_0|) e^{-|p_0|r}
\end {equation}
is replaced by
\begin {equation}
   \sum_{\{q_0\}} e^{-|q_0|r} e^{-|p_0+q_0|r}
      = (\csch\, \bar r + |\bar p_0|) e^{-|p_0|r} \,,
\end {equation}
where $p_0$ represents bosonic frequencies and
\begin {equation}
   \bar r \equiv 2\pi T r,
   \qquad
   \bar p_0 \equiv p_0/2\pi T \,.
\end {equation}
This has the effect of simply replacing occurrences of $\coth$ (and its small
$r$ expansion) in the bosonic derivation by $\csch$ (and its small $r$
expansion).  So, for instance,
\begin {equation}
   \PifT(P) =
     {T\over(4 \pi)^2} \int d^3 r \, {1\over r^2}
           e^{i \vec p \cdot \vec r}
           \left(\csch\, \bar r - {1\over\bar r}\right)
           e^{-|p_0|r} + O(\eps)\,,
\label {PifT integral}
\end {equation}
where $\PifT$ is the finite-temperature contribution to $\Pif$.
Similarly,
\begin {eqnarray}
   \sumint_P && \left\{ [\PifT(P)]^2
        - \left({2\over P^2}\sumint_{\{Q\}}{1\over Q^2}\right)^2 \right\}
\nonumber\\
   && \qquad\qquad
   = {T^4 \over 32\pi^2} \int\nolimits_0^\infty d\bar r \, \bar r^{-2}
               \left[ \left(\csch \bar r - {1\over\bar r}\right)^2
                          - \left(-\bar r\over 6\right)^2 \right]
               (\coth \bar r - 1)
\nonumber \\
   && \qquad\qquad\qquad
   + {T^4\over32\pi^2} \int\nolimits_0^\infty d\bar r \, \bar r^{-2}
      \left( \csch \bar r - {1\over\bar r} \right)^2
     + O(\eps)
\nonumber \\
   && \qquad\qquad
   = {1\over(4\pi)^2} \left(T^2\over12\right)^2 \left[
        2 {\zeta'(-3)\over\zeta(-3)}
        - 2\gammaE + {73\over 15}
        - {6\over5} \ln 2
   \right]
   + O(\eps) \,.
\end {eqnarray}
The integrals in the last equation were performed using the method
of Appendix~\ref{integration appendix}.  Putting the above result together with
(\ref{b1 eqn}, \ref{b2 eqn}, \ref{f eqn}) gives
\begin {equation}
   \sumint [\PifT]^2
   = {1\over(4\pi)^2} \left(T^2\over12\right)^2 \left[
        {1\over\eps} + 6 \lnmub
        + 2 {\zeta'(-3)\over\zeta(-3)}
        + 4 {\zeta'(-1)\over\zeta(-1)}
        + {133\over15}
        - {26\over5} \ln 2 \right]
   + O(\eps) \,.
\label {PifT2 integral}
\end {equation}

The remaining terms needed to evaluate $I_\bball^\xff$ are discussed in
Appendix~\ref{bball appendix}.
The final result for the fermionic basketball is
\begin {equation}
   I_\bball^\xff
   = {1\over(4\pi)^2} \left(T^2\over12\right)^2 \left[
        {3\over2\eps} + 9 \lnmub
        - 3 {\zeta'(-3)\over\zeta(-3)}
        + 12 {\zeta'(-1)\over\zeta(-1)}
        + {173\over20} - {63\over5} \ln 2 \right]
   + O(\eps) \,.
\label{ffball result}
\end {equation}
This agrees with the numerical result of ref.~\cite{Coriano&Parwani}.

% --------------------------------------------------------------------------

\subsection{The mixed basketball $I_\bball^\xbf$}

As has been noted by Parwani and Corian\`o\cite{Coriano&Parwani},
$I_\bball^\xbf$ can be written
in terms of $I_\bball^\xbb$ and $I_\bball^\xff$ by the trick of
rescaling momenta $(p_0,\vec p)$ by a factor of 2:
\begin {eqnarray}
   I_\bball^\xbb
   &=& \sumint_{PQKR} {\delta(P+Q+K+R) \over P^2 Q^2 K^2 R^2}
\nonumber \\
   &=& 2^{3d-11} \sumint_{P+\{P\}} \sumint_{Q+\{Q\}}
                 \sumint_{K+\{K\}} \sumint_{R+\{R\}}
                      {\delta(P+Q+K+R) \over P^2 Q^2 K^2 R^2}
\nonumber \\
   &=& 2^{3d-11} \left[ I_\bball^\xbb + 6 I_\bball^\xbf + I_\bball^\xff
                 \right] \,,
\end {eqnarray}
so that
\begin {equation}
   I_\bball^\xbf = - {1\over6} \left(1-2^{11-3d}\right) I_\bball^\xbb
                   - {1\over6} I_\bball^\xff \,.
\end {equation}
To simplify notation above, we have used the short-hand
\begin{equation}
    \delta (P+Q+K+R) \equiv \mu^{-2 \epsilon} {1 \over T}
	\delta_{p_0{+}q_0{+}k_0{+}r_0}
        (2 \pi)^{d-1}
        \delta^{(d-1)} (\vec p{+} \vec q{+}\vec k{+}\vec r) \,.
\end{equation}

% --------------------------------------------------------------------------

\subsection{The new integral $H_3$}

We now turn to the integral $H_3$ of (\ref{H3 def}), which is the one integral
that is not directly analogous to a previous bosonic calculation.
However, our attack on $H_3$ is inspired by our derivation of
$I_\qcd^\xbb \sim \textsumint P^{-4} (\Delta\Pi^\b_{\mu\nu})^2$
in ref.~\cite{Arnold&Zhai}, where we noted that the orthogonality of
$\Pi^\b_{\mu\nu}$ to $P_\mu$ lead to useful algebraic simplifications.
We will therefore rewrite $H_2$ in an analogous form.  First note that
$H_3$ is of the form
\begin {equation}
   H_3 = \sumint_{\{P\}} {1\over P^2} \left[ A_\mu(P) \right]^2 \,,
\end {equation}
where
\begin {equation}
   A_\mu (P) \equiv \sumint_Q {Q_\mu \over Q^2 (P+Q)^2} \,.
\end {equation}
Our method is to replace $A_\mu$ by something that is orthogonal to $P_\mu$.
So define
\begin {equation}
   I_3 \equiv \sumint_{\{P\}} {1\over P^2} \left[ J_\mu(P) \right]^2 \,,
\end {equation}
where
\begin {equation}
   J_\mu (P) \equiv \sumint_Q {(2Q+P)_\mu \over Q^2 (P+Q)^2}
      - {P_\mu\over P^2} \left(
          \sumint_Q {1\over Q^2} - \sumint_{\{Q\}} {1\over Q^2} \right) \,.
\end {equation}
It is easy to verify using our standard reduction tricks that
$P \cdot J {=} 0$ and that
\begin {equation}
   H_3 = {1\over4} I_3 + {1\over4} I_\bball^\xbf + {1\over4}(b_1-f_1)^2 f_2
            - {1 \over 2} (b_1-f_1) I_\sun^\f \,.
\label {H3 vs I3}
\end {equation}

Now focus on $I_3$.
The orthogonality of $J$ to $P$ and Lorentz invariance imply that
$J_\mu$ has the form
\begin {equation}
   J_\mu(P) =
      \left( n_\mu - {n\cdot P\over P^2} P_\mu \right) f(P) \,,
\end {equation}
where $n_\mu = (1, \vec 0)$ is the four-velocity of the thermal bath.
Then
\begin {equation}
   J_\mu(P) = {P^2\over p^2} \left( n_\mu - {p_0\over P^2} P_\mu \right)
      j_0(P)
\end {equation}
and
\begin {equation}
   I_3 = \sumint_{\{P\}} {1\over p^2} \left[j_0(P)\right]^2 \,.
\label {useful I3 eqn}
\end {equation}
The next simplification occurs by noting that $J_\mu$ vanishes at
zero temperature because
\begin {equation}
   \int {d^d Q \over (2\pi)^d} {(2Q+P)_\mu \over Q^2(P+Q)^2} = 0
\end {equation}
by anti-symmetry under $Q \to {-}(Q{+}P)$.  The large $P$ behavior of
$j_0(P)$ is therefore the large $P$ behavior of its finite-temperature
piece $j^{(T)}_0(P)$, which is $O(1/P^3)$ because the $O(1/P)$ behavior
of the individual pieces cancels:
\begin {equation}
   \left(\sumint_{\{P\}} {(2Q+P)_\mu \over Q^2(P+Q)^2}\right)^{(T)}
   \longrightarrow {P_\mu\over P^2} \left(
          \sumint_Q {1\over Q^2} - \sumint_{\{Q\}} {1\over Q^2} \right)
	\qquad \hbox{as fermionic $P \to \infty$.}
\end {equation}
As a result, $I_3$ is both UV and IR finite as $\eps{\to}0$.  So we can set
$d=4$ and evaluate $j_0$ for fermionic $p_0$ in terms of the massless
scalar propagator $\Delta$:
\begin {eqnarray}
   j_0(P)
   &=& T \sum_{q_0} \int d^3r \, e^{i\vec p\cdot\vec r}
          \Delta(q_0,\vec r) \, \Delta(p_0+q_0,\vec r)
          \, (2q_0+p_0)
       \, - \, {T^2\over 8} {p_0\over P^2} + O(\eps)
\nonumber \\
   &=& {T^2\over (4\pi)^2} \sum_{q_0} \int d^3r \,
          {1\over r^2} e^{i\vec p\cdot\vec r}
          e^{-|q_0|r} e^{-|p_0+q_0|r} (2q_0+p_0)
        \, - \, {T^2\over 8} {p_0\over P^2} + O(\eps)
\nonumber \\
   &=& {T^2\over 16\pi} \int d^3r \,
          {1\over r^2} e^{i\vec p\cdot\vec r}
          \left\{
             \partial_{\bar r} \left[
                 (\csch\bar r - \coth\bar r) e^{-|p_0|r} \right]
             + {1\over2} e^{-|p_0|r}
          \right\} \sgn p_0
          - {T^2\over8} {p_0\over P^2} + O(\eps)
\nonumber \\
   &=& - {T\over8\pi} \int\nolimits_0^\infty dr \,
         \left(\partial_r {\sin pr \over pr} \right)
         \left(\csch\bar r - \coth\bar r + \textstyle{1\over2}\bar r\right)
         e^{-|p_0|r} \sgn p_0  + O(\eps) \,,
\end {eqnarray}
where $\sgn p_0$ means the sign ($\pm 1$) of $p_0$.
Plugging into (\ref{useful I3 eqn}), doing the fermionic $p_0$ sum, and
using the identity
\begin {equation}
   {1\over(2\pi)^3} \int {d^3p\over p^2} \,
       {\sin pr \over pr} \, {\sin ps\over ps}
   = {1\over 4\pi} \left[ {1\over r}\theta(r-s) + {1\over s}\theta(s-r)
                        \right]
\end {equation}
to do the $p$ integral then yields
\begin {eqnarray}
   I_3 &=& {T^4\over 128\pi^2} \int\nolimits_0^\infty d\bar r\, \bar r^{-2}
      \left(\csch\bar r - \coth\bar r + \textstyle{1\over2}\bar r\right)^2
      \csch\bar r + O(\eps)
\nonumber \\
   &=& {1\over(4\pi)^2} \left(T^2\over12\right)^2 \left[
        {9\over2} {\zeta'(-3)\over\zeta(-3)}
        - 9 {\zeta'(-1)\over\zeta(-1)}
        + {9\over2} \gammaE + {9\over2} + {117\over10} \ln 2 \right]
   + O(\eps) \,.
\end {eqnarray}
Again, we have used the methods of Appendix~\ref{integration appendix}
to do the integrals.
Using (\ref{H3 vs I3}) to relate $I_3$ to $H_3$ finally gives
\begin {equation}
   H_3 = {1\over(4\pi)^2} \left(T^2\over12\right)^2 \left[
        {3\over8\eps} + {9\over4} \lnmub
        + {3\over2} {\zeta'(-3)\over\zeta(-3)}
        - {3\over2} {\zeta'(-1)\over\zeta(-1)}
        + {9\over4} \gammaE + {361\over160} + {57\over10} \ln 2 \right]
   + O(\eps) \,.
\end {equation}
This agrees, within errors, with the numerical results
of ref.~\cite{Coriano&Parwani}.%
\footnote{
   We have also made a more precise numerical test of our analytic methods
   by computing (\ref{useful I3 eqn}) by brute force: we did the $p$ integral
   and Euclidean $p_0$ sum numerically and used the contour trick to
   get an integral form for $j_0(p_0,p)$, which we also evaluated numerically.
   The $p_0$ sum converges quite quickly, and summing
   $\bar p_0$ up to $\pm {5/2}$ gave agreement with out analytic result
   to 0.05\%.
}

% ==========================================================================

\section {Results and Discussion}

The evaluation of the final basic integrals $I_\qcd^\xff$ and
$I_\qcd^\xbf$ closely parallel the derivation of $I_\qcd^\xbb$ in
ref.~\cite{Arnold&Zhai},
and we leave the details for Appendix~\ref{qcd appendix}.
Combining all the results for individual graphs collected in
Appendix~\ref{graphs appendix},
our final result for the free energy is
\begin{eqnarray}
    F &=& \da T^4 {\pi^2\over9} \Biggr \{
         - {1 \over 5} \left(1 + {7\df\over4\da} \right)
	 + \left ({g \over 4 \pi} \right )^2
              \left(\ca+\textstyle{5\over2}\sf\right)
\nonumber\\
	&& \qquad
	 - {16 \over \sqrt{3}} \left ({g \over 4 \pi} \right )^3
              (\ca+\sf)^{3\over2}
	 - 48 \left ({g \over 4 \pi} \right )^4 \ca (\ca+\sf)
              \ln\left({g\over 2\pi}\sqrt{\ca+\sf \over 3}\right)
\nonumber\\
	&& \qquad
        + \left(g \over 4 \pi\right )^4 \ca^2 \Biggr [
	    {22 \over 3} \lnmub
	    {+}{38 \over 3} {\zeta'(-3) \over \zeta (-3)}
	    {-}{148 \over 3} {\zeta'(-1) \over \zeta (-1)}
	    {-}4 \gammaE
	    {+}{64 \over 5}
          \Biggr ]
\nonumber\\
	&& \qquad
        + \left(g \over 4 \pi\right )^4 \ca\sf \Biggr [
	    {47 \over 3} \lnmub
	    {+}{1 \over 3} {\zeta'(-3) \over \zeta (-3)}
	    {-}{74 \over 3} {\zeta'(-1) \over \zeta (-1)}
	    {-}8 \gammaE
	    {+}{1759 \over 60}
            {+}{37\over5}\ln 2
          \Biggr ]
\nonumber\\
	&& \qquad
        + \left(g \over 4 \pi\right )^4 \sf^2 \Biggr [
	    {-}{20 \over 3} \lnmub
	    {+}{8 \over 3} {\zeta'(-3) \over \zeta (-3)}
	    {-}{16 \over 3} {\zeta'(-1) \over \zeta (-1)}
	    {-}4 \gammaE
	    {-}{1 \over 3}
            {+}{88\over5}\ln 2
          \Biggr ]
\nonumber\\
	&& \qquad
        + \left(g \over 4 \pi\right )^4 \stf \Biggr [
	    {-}{105 \over 4}
            {+}24\ln 2
          \Biggr ]
	+ O(g^5) \Biggr \} \,.
\end{eqnarray}
Evaluated numerically for QCD with $\nf$ quark flavors, this is
\begin {eqnarray}
   F = - && {8 \pi^2 T^4\over 45} \biggl\{
     1 + \textstyle{21\over32}\nf
     - 0.09499\, g^2 \left(1 + \textstyle{5\over12}\nf\right)
     + 0.12094\, g^3 \left(1+\textstyle{1\over6}\nf\right)^{3/2}
\nonumber\\
&&
     + g^4 \biggl[
         0.08662 \left(1+\textstyle{1\over6}\nf\right)
                     \ln\left(g\sqrt{1+\textstyle{1\over6}\nf}\right)
         - 0.01323 \left(1+\textstyle{5\over12}\nf\right)
                   \left(1-\textstyle{2\over33}\nf\right)
                   \ln{\bar\mu\over T}
\nonumber\\
&& \qquad
         + 0.01733 - 0.00763\,\nf - 0.00088\,\nf^2
     \biggr]
     + O(g^5) \biggr\}\,.
\end {eqnarray}
For QED with $\nf$ massless charged fermions with charges $q_i e$,
the free energy is
\begin {eqnarray}
   F = - {\pi^2 T^4\over 45} && \biggl\{
     1 + \textstyle{7\over4}\nf
     - 0.07916\, e^2 \sum q_i^2
     + 0.02328\, e^3 \left(\sum q_i^2\right)^{3/2}
\nonumber \\
&& \quad
     + e^4 \left[
         \left(-0.00352 + 0.00134\ln{\bar\mu\over T}\right)
               \left(\textstyle\sum q_i^2\right)^2
         + 0.00193 \textstyle\sum q_i^4
     \right]
     + O(e^5) \biggr\}\,.
\end {eqnarray}
Our QED result agrees, within errors, with the purely numerical derivation
of ref.~\cite{Coriano&Parwani}.
Ref.~\cite{Coriano&Parwani} also gives the QED result for the $O(e^5)$ piece.

\begin {figure}
\vbox
    {%
    \begin {center}
	\leavevmode
	
	\epsfbox [150 250 500 550] {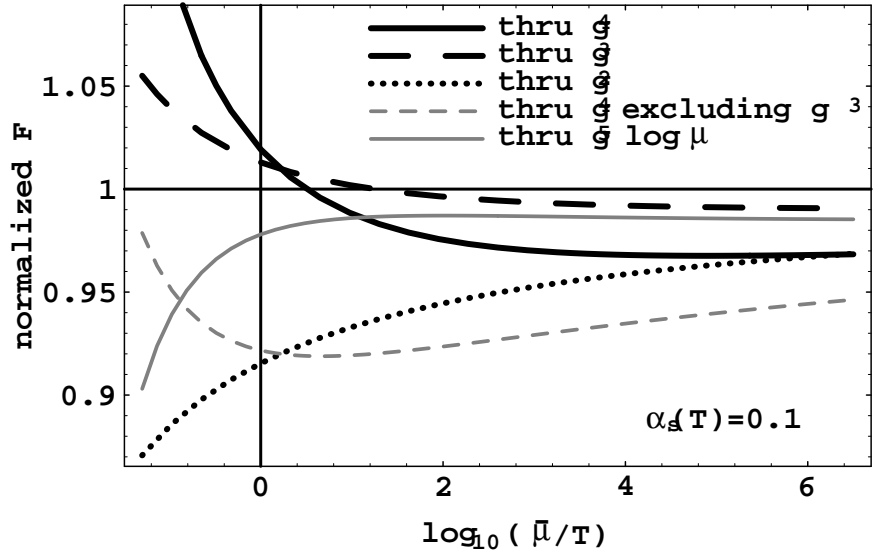}
    \end {center}
    \caption
	{%
	\label {figfh1}
        The dependence of the free energy $F$ on the choice of renormalization
        scale $\bar\mu$ for 6-flavor QCD with $\alphas(T) = 0.1$.
        The free energy is normalized in units of the ideal gas result
        $-({1\over45}\da + {7\over180}\df)  \pi^2 T^4$.
        The thick solid, dashed, and dotted lines are the results for $F$
        including terms through $g^4$, $g^3$, and $g^2$ respectively.
        The light solid curve is the $g^4$ result plus the
        $g^5 \ln(\bar\mu/T)$ term required by renormalization group
        invariance.  The light dashed curve is the $g^4$ result minus the
        $g^3$ term.
	}%
    }%
\end {figure}

As in ref.~\cite{Arnold&Zhai}, we can now investigate whether the
perturbative expansion of the QCD free energy is well-behaved
for physically-realized values of couplings.
Fig.~\ref{figfh1} shows the result for 6-flavor QCD when $\alphas(T){=}0.1$
(which corresponds to scales of order a few 100 GeV).  The free energy
is plotted vs.\ the choice of renormalization scale $\bar\mu$.
We have taken
\begin {equation}
   {1\over g^2(\bar\mu)} \approx
      {1\over g^2(T)} - \beta_0 \ln{\bar\mu\over T}
      + {\beta_1\over\beta_0}
         \ln\left(1 - \beta_0 g^2(T) \ln{\bar\mu\over T}\right)
      \,,
\end {equation}
where
\begin{equation}
   \beta_0 = {1\over(4 \pi)^2}
     \left(-\textstyle{22\over3}\ca + \textstyle{8\over3}\sf\right) \,,
   \qquad
   \beta_1 = {1\over(4 \pi)^4}
     \left(-\textstyle{68\over3}\ca^2 + \textstyle{40\over3}\ca\sf
           + 8\stf \right) \,.
\end {equation}

\begin {figure}
\vbox
    {%
    \begin {center}
	\leavevmode
	
	\epsfbox [150 250 500 550] {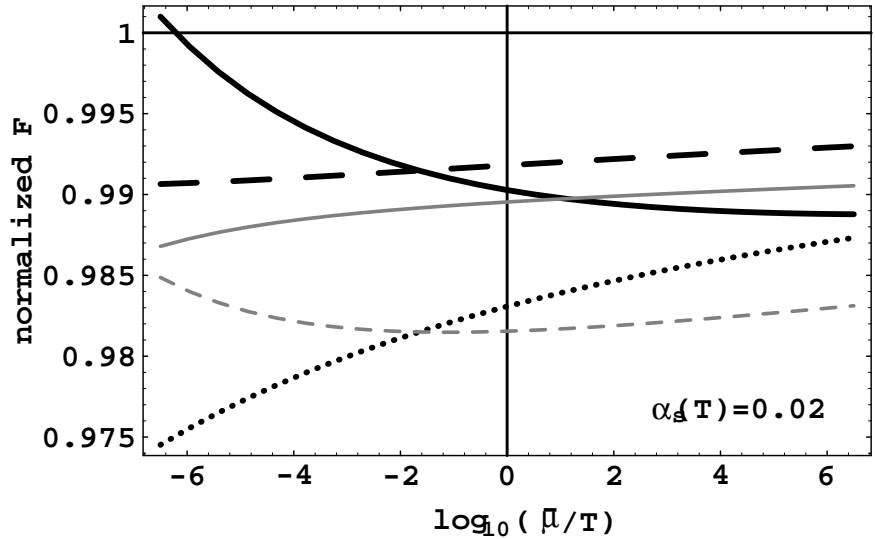}
    \end {center}
    \caption
	{%
	\label {figfh2}
        The same as fig.~\protect\ref{figfh1} but for $\alphas(T) = 0.02$.
	}%
    }%
\end {figure}

\begin {figure}
\vbox
    {%
    \begin {center}
	\leavevmode
	
	\epsfbox [150 250 500 550] {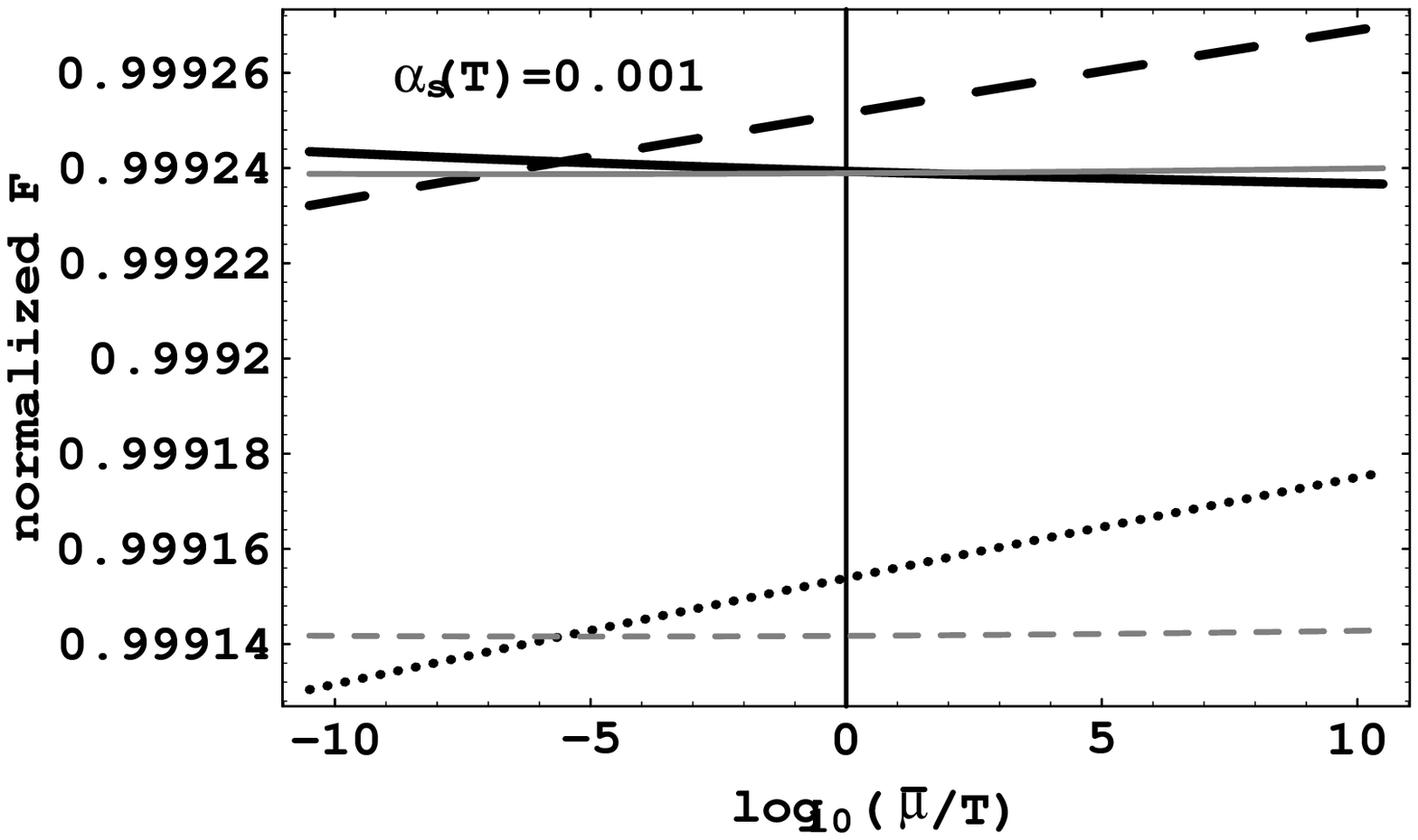}
    \end {center}
    \caption
	{%
	\label {figfh3}
        The same as fig.~\protect\ref{figfh1} but for $\alphas(T) = 0.001$.
	}%
    }%
\end {figure}

\begin {figure}
\vbox
    {%
    \begin {center}
	\leavevmode
	
	\epsfbox [150 250 500 550] {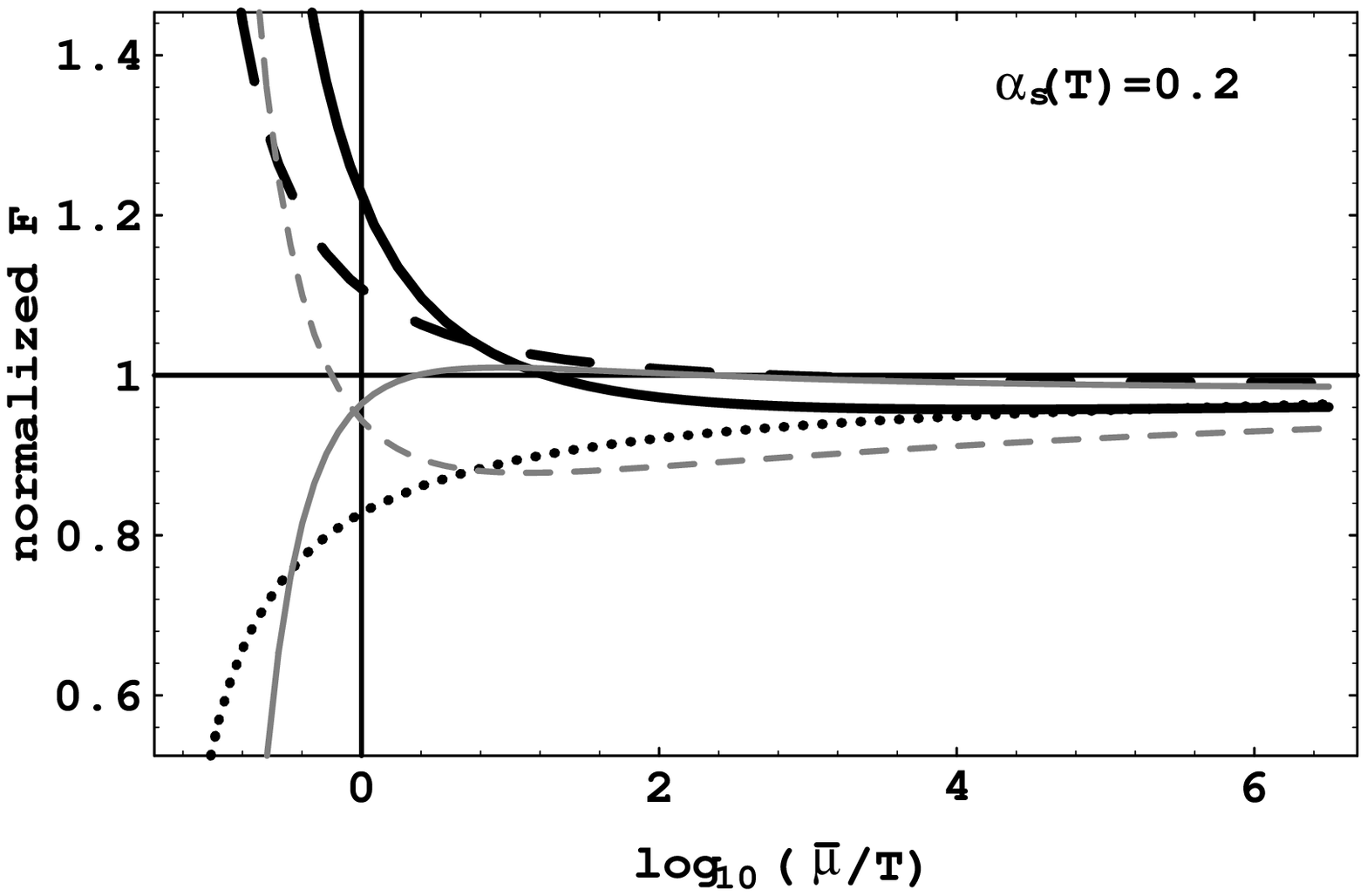}
    \end {center}
    \caption
	{%
	\label {figfh0}
        The same as fig.~\protect\ref{figfh1} but for $\alphas(T) = 0.2$
        and $\nf = 5$.
	}%
    }%
\end {figure}

If the expansion is well-behaved, the result for $F$ should become more
independent of $\bar\mu$ as higher-order corrections are included.
Instead, we see that it does not.  In ref.~\cite{Arnold&Zhai}, we
argued that the $g^3$ term involves different
physics than the $g^2$ term, and that it should perhaps be treated
separately when discussing the behavior of the series.  Ideally, one
should calculate the free energy through $g^5$, which is the first
order that compensates for the $\mu$ dependence of the $g^3$ term.
With our present results, however, we can at least follow
ref.~\cite{Arnold&Zhai} and plot results where (1) we artificially exclude the
$g^3$ term, or (2) we improve our $O(g^4)$ result with the $g^5 \ln\mu$
term required by renormalization-group invariance:
\begin {equation}
    \Delta F =
    \da T^4 {\pi^2\over9} \Biggr \{
        {-}{16 \over \sqrt{3}} \left(g \over 4\pi\right)^5
            (\ca+\sf)^{3\over2} (11 \ca - 4 \sf) \lnmub
    \Biggl\} \,.
\end {equation}
Either modification can be seen to somewhat improve the behavior of
the perturbative expansion of fig.~\ref{figfh1}.  For comparison with our
previous results for the pure gauge case in ref.~\cite{Arnold&Zhai},
figs.~\ref{figfh2} and \ref{figfh3} show similar plots for the
smaller couplings
$\alphas(T){=}0.02$ and $\alphas(T){=}0.001$, where the perturbative
expansion becomes progressively better behaved, as it should.
For those readers who might be interested in the behavior of the
expansion at scales of order several GeV, fig.~\ref{figfh0} shows our results
for the case $\alphas(T)=0.2$.

%%%%%%%%%%%%%%%%%%%%%%%%%%%%%%%%%%%%%%%%%%%%%%%%%%%%%%%%%%%%%%%%%%%%%%%%%%

\bigskip

This work was supported by the U.S. Department of Energy,
grant DE-FG06-91ER40614.  C.Z. was also partly supported by
U.S. Department of Energy grant DE-AC02-76ER01428.

%%%%%%%%%%%%%%%%%%%%%%%%%%%%%%%%%%%%%%%%%%%%%%%%%%%%%%%%%%%%%%%%%%%%%%%%%%
%%%%%%%%%%%%%%%%%%%%%%%%%%%%%%%%%%%%%%%%%%%%%%%%%%%%%%%%%%%%%%%%%%%%%%%%%%

\newpage
\appendix

\section{Results for individual graphs}
\label {graphs appendix}

Writing F=$\mu^{-2\eps}\da {\cal F}$ and ignoring terms of $O(\eps)$, the
diagrams of fig.~\ref{figf} are given in Appendix A
of ref.~\cite{Arnold&Zhai} except that
\begin {mathletters}%
\label {gauge diagrams}%
\begin {eqnarray}
\setcounter{eqletter}{3}    % I want the label to be (c)
   -{\cal F}^{\rm c} &=&
      -{1\over8\pi} (M_1^2 + M_2^2 + M_3^2 + M_4^2) M T
   \,,
\\
\setcounter{eqletter}{12}    % I want the label to be (l)
   -{\cal F}^{\rm l} &=&
      {1\over4} g^4 \left[
         \ca^2 I_\qcd^\xbb + 2 \ca\sf I_\qcd^\xbf + \sf^2 I_\qcd^\xff \right]
      + O(g^5)
   \,.
\\
\noalign{ \hbox{The diagrams of fig.~\ref{figf2} are given by} }
\nonumber\\
\setcounter{eqletter}{14}    % I want the labels to start with (n)
   -{\cal F}^{\rm n} &=&
      {2\df\over\da} \sumint_{\{P\}} \ln P^2
   \,,
\\
   -{\cal F}^{\rm o} &=&
      \sf g^2 Z_g^2 \left[ (d-2) f_1 (2b_1-f_1) + \delta_3 \right]
   \,,
\\
   && \delta_3 =
       M^2 \sumint_P {\delta_{p_0^{}} \over P^2} \Pif(P)
       - 4 I^\f_\resum
       + {1\over8\pi} {M_4^2 M T\over \sf g^2 Z_g^2}
   \,,
\nonumber
\\
   -{\cal F}^{\rm p} &=&
      \left( -\textstyle{1\over2}\sf \ca + \stf \right)
      g^4 (d-2) \left[ {(6-d)\over2} I_\bball^\xff - 2\eps I_\bball^\xbf
                \right]
   \,,
\\
   -{\cal F}^{\rm q} &=&
      \stf g^4 (d-2)^2
      \left[ - I_\bball^\xbf + 2 H_3 - f_2 (f_1-b_1)^2 \right]
   \,,
\\
   -{\cal F}^{\rm r} &=&
      \ca \sf g^4 (d-2) I_\bball^\xbf
   \,.
\end {eqnarray}
\end {mathletters}
The resummation of the gauge boson line in diagram (o) is similar to the
resummation of diagram (e), discussed in ref.~\cite{Arnold&Zhai}.
In particular,
\begin {equation}
   I^\f_\resum
   \equiv \sumint_{P\{Q\}} \left[  {\delta_{p_0^{}} \over P^2+M^2}
                     - {\delta_{p_0^{}} \over P^2}     \right]
     \left[ {q_0^2 \over Q^2 (P+Q)^2} - {q_0^2 \over Q^4} \right]
   = O(g^3,\eps) \,.
\end {equation}
The last equality follows from a derivation similar to the bosonic case
treated in ref.~\cite{Arnold&Zhai} and means that $I^\f_\resum$ can
be ignored at the order under consideration.

The multiplicative renormalization constant used for the coupling is
given by
\begin {equation}
    g_\bare = Z_g g \mu^\epsilon
     = \left[ 1 - {11\over6} {\ca g^2 \over (4\pi)^2\eps}
                + {2\over3} {\sf g^2 \over (4\pi)^2\eps} + O(g^4) \right]
	g \mu^\epsilon \,.
\end {equation}
The vector mass $M$ is given by (\ref{M def}), and $M_4^2$ is the piece
of $M^2$ due to the fermion contribution of fig.~\ref{fig pi}(b).

% ==========================================================================

\section{Large $P$ behavior of $\Pif(P)$}
\label {pi appendix}

In Appendix B of ref.~\cite{Arnold&Zhai}, we derived the large $P$ behavior of
the bosonic $\PibT(P)$ to be
\begin {equation}
    \PibT(P) = 2 J^\b_{-1} {T^2 \over P^2} + 8 J^\b_1 {T^4 \over P^6}
	\left ({p^2 \over d-1} - p_0^2 \right )
	+ O (T^6 / P^6) \,,
\label{PibT_exp_result}
\end{equation}
where
\begin {equation}
   J^\b_\alpha
	= \left ({4 \pi \mu^2 \over T^2} \right )^\epsilon
	{\Gamma (3{-}2 \epsilon{+}\alpha) \zeta (3{-}2 \epsilon{+}\alpha)
	\over 4 \pi^{3/2}
             \Gamma\left( {\textstyle{3\over2}} - \epsilon \right)  }
        \,.
\end {equation}
The easiest way to get a similar expression for the fermionic $\PifT$ is
to note that, by scaling momenta $(q_0,q)$ by a factor of 2, one gets
\begin {equation}
   \sumint_{Q+\{Q\}} {1\over Q^2(P+Q)^2}
   = 2^{5-d} \sumint_{Q} {1\over Q^2 (2P+Q)^2} \,,
\end {equation}
and so
\begin {equation}
   \Pif(P) = 2^{5-d} \Pib(2P) - \Pib(P) \,.
\label {Pif Pib relation}
\end {equation}
Applying this identity to (\ref{PibT_exp_result}) then gives the same
formula,
\begin {equation}
    \PifT(P) = 2 J^\f_{-1} {T^2 \over P^2} + 8 J^\f_1 {T^4 \over P^6}
	\left ({p^2 \over d-1} - p_0^2 \right )
	+ O (T^6 / P^6) \,,
\label{PifT_exp_result}
\end{equation}
for the fermionic case but with
\begin {equation}
   J^\f_\alpha = \left(2^{2-d-\alpha}-1 \right) J^\b_\alpha \,.
\end {equation}

% ==========================================================================

\section{New integrals of hyperbolic functions}
\label {integration appendix}

In Appendix C of ref.~\cite{Arnold&Zhai},
we showed how to evaluate convergent integrals
of the form
\begin {equation}
    I = \int\nolimits_0^\infty\! dx \left(
           \sum_{m,n} c_{mn} x^m \coth^n x
           + \sum_m d_m x^m e^{-a_m x}
        \right)
\end {equation}
by regulating the individual terms by introducing an extra factor of
$x^\delta$ and taking $\delta{\to}0$ at the end.
In the current work, we need to extend our set of integrals to
\begin {equation}
    I = \int\nolimits_0^\infty\! dx \left(
           \sum_{m,n,p} c_{mnp} x^m \coth^n x \, \csch^p x
           + \sum_m d_m x^m e^{-a_m x}
        \right) \,.
\end {equation}
To handle this, we need in addition to our previous basic regulated integrals
\begin {eqnarray}
   \int\nolimits_0^\infty\! dx\, x^z &=& 0 \,,
\\
   \int\nolimits_0^\infty\! dx\, x^z \coth x
   &=& 2^{-z} \Gamma(z+1) \, \zeta(z+1) \,,
\\
   \int\nolimits_0^\infty\! dx\, x^z e^{-ax}
   &=& a^{-1-z} \Gamma(1+z) \,,
\\
\noalign {\hbox{the new integral}}
\nonumber\\
   \int\nolimits_0^\infty\! dx\, x^z \csch x
   &=& (2-2^{-z}) \Gamma(z+1) \, \zeta(z+1) \,.
\end {eqnarray}
In addition, we need to generalize our previous recursion relation to
\begin {eqnarray}
   \int\nolimits_0^\infty\! dx\, && x^z \coth^n x \, \csch^p x =
\nonumber \\
   &&
   \int\nolimits_0^\infty\! dx\, \left[
          {z\over p+n-1} x^{z-1} \coth^{n-1} x \, \csch^p x
          + {n-1\over p+n-1} x^z \coth^{n-2} x \, \csch^p x
       \right]
\end {eqnarray}
and also use
\begin {equation}
   \int\nolimits_0^\infty\! dx\, x^z \coth^n x \, \csch^p x
   = \int\nolimits_0^\infty\! dx\, \left[
          x^z \coth^{n+2} x \, \csch^{p-2} x
          - x^z \coth^n x \, \csch^{p-2} x
       \right] \,.
\end {equation}

% ==========================================================================

\section {Completion of the calculation of $I_\bball^\xff$}
\label {bball appendix}

In this section, we complete the evaluation of the fermionic basketball
integral.  The derivation directly parallels the bosonic case treated in
Appendix D of ref.~\cite{Arnold&Zhai},
with hyperbolic cotangents becoming hyperbolic
cosecants as we discussed earlier.  We therefore refer the reader to
ref.~\cite{Arnold&Zhai}
and shall here simply present the differences for a selected few
intermediate results.

As in the bosonic case, we write
\begin {equation}
   \sumint_P \PifT \PiZ = I_{\rm a}^\f + I_{\rm b}^\f + I_{\rm c}^\f \,,
\end {equation}
where
\begin{equation}
    I^\f_{\rm a} \equiv \sumint_P \left [\Pi^{(0)} (P)
	- {1 \over (4 \pi)^2\epsilon} \right ]
	 \left [\PifT (P)
	 - \PifT_{\rm UV} (P) \right ] \,,
\label{I_a_def}
\end{equation}
\begin{equation}
    I^\f_{\rm b} \equiv {1 \over (4 \pi)^2\eps} \,\,
	\sumint_P \left [\PifT (P)
	 - \PifT_{\rm UV} (P) \right ] \,,
\label{I_b_def}
\end{equation}
\begin{equation}
    I^\f_{\rm c} \equiv \sumint_P \Pi^{(0)} (P)
	 \PifT_{\rm UV} (P) \,,
\label{I_c_def}
\end{equation}
and
\begin{equation}
    \PifT_{\rm UV} (P) \equiv
	2 J^\f_{-1} {T^2 \over P^2} + (1 - \delta_{p_0^{}})
	8 J^\f_1 {T^4 \over P^6} \left ({p^2 \over d-1} - p_0^2 \right ) \,.
\label{PiT_UV_def}
\end{equation}
The zero-temperature piece $\PiZ$ of $\Pif$ is the same as the bosonic
case.
One finds
\begin{eqnarray}
    I^\f_{\rm a} &=& {T^4 \over (4 \pi)^2} {1\over2}
        \int_0^{\infty} {d \bar r \over \bar r^3} \Biggr \{
        \left (\csch\, \bar r - {1 \over \bar r} + {\bar r \over 6}
           - {7\bar r^3 \over 360} \right )
        \left (1 - {\bar r \over 2} {d \over d\bar r} \right )
           (\coth\bar r - 1)
\nonumber \\
&& \qquad\qquad\qquad\qquad
        + \left (\csch\, \bar r - {1 \over \bar r}
              + {\bar r \over 6} \right ) \Biggr \}
        + O(\eps)
\nonumber \\
    &=& {1 \over (4 \pi)^2}
        \left ({T^2 \over 12} \right )^2
        \left [
	  - {29 \over 10} {\zeta'(-3) \over \zeta (-3)}
          + 5 {\zeta'(-1) \over \zeta (-1)}
	  - {21 \over 10} \gammaE
          + {43 \over 30}
          - {27\over10} \ln 2
        \right ]
    + O(\eps) \,,
\label{If_a_result}
\\
   I^\f_{\rm b} &=&
	{T^4 \over (4 \pi)^2} {1 \over \epsilon}
	\left \{ \left(J^\f_{-1} - 2 J^\b_{-1}\right) J^\f_{-1}
	- A \, \left ({4 \pi \mu^2 \over T^2} \right )^{\epsilon}
	S_0 (\epsilon) - {8 J^\f_1 \over d-1} \left [
	S_0 (2) - d \, S_1 (3) \right ] \right \}
\nonumber \\
    &=& {1 \over (4 \pi)^2}
        \left ({T^2 \over 12} \right )^2
        \left [
	  - {29 \over 10} {\zeta'(-3) \over \zeta (-3)}
          + 5 {\zeta'(-1) \over \zeta (-1)}
	  - {21 \over 10} \gammaE
          + {43 \over 30}
          - {27\over10} \ln 2
        \right ]
    + O(\eps) \,,
\label {If_b_result}
\\
    I^\f_{\rm c} &=& 2 A \, T^4 \,
	\left ({4 \pi \mu^2 \over T^2} \right )^\epsilon
	\left \{ J^\f_{-1} S_0 (1{+}\epsilon) +
        {4 J^\f_1 \over d{-}1} \left [S_0 (2{+}\epsilon)
	- d \, S_1 (3{+}\epsilon) \right ] \right \}
\nonumber\\
	&=&  {1 \over (4 \pi)^2}
	\left ({T^2 \over 12} \right )^2
	\left [{1 \over 20\epsilon}
	+ {3 \over 10} \ln {{\bar \mu} \over 4 \pi T}
	+ {21 \over 10} {\zeta'(-3) \over \zeta(-3)}
	- 6 {\zeta'(-1) \over \zeta(-1)}
	+ {21 \over 5} \gammaE
	- {43\over8}
        + {17\over10} \ln 2    \right ]
\nonumber \\
&& \qquad\qquad\qquad\qquad
   + O (\epsilon) \,,
\label{If_c_result}
\end{eqnarray}
where $A$ and $S_n(\alpha)$ are defined in ref.~\cite{Arnold&Zhai}.
Putting together (\ref{If_a_result}), (\ref{If_b_result}),
and (\ref{If_c_result}),
\begin{eqnarray}
    \sumint \Pi^{f(T)} \Pi^{(0)}
	= {1 \over (4 \pi)^2}
	\left ({T^2 \over 12} \right )^2
&&
	\left [{1 \over 20 \epsilon}
	+ {3 \over 10} \ln {{\bar \mu} \over 4 \pi T}
	- {37 \over 10} {\zeta'(-3) \over \zeta(-3)}
	+ 4 {\zeta'(-1) \over \zeta(-1)}
	- {301 \over 120}
        - {37 \over 10} \ln 2  \right ]
\nonumber \\
&& \qquad
    + O(\eps) \,.
\label{secondterm}
\end {eqnarray}
The result for $\textsumint \left[\PiZ\right]^2$ is given
in ref.~\cite{Arnold&Zhai},
and adding it to (\ref{PifT2 integral}) and (\ref{secondterm}) yields the final
result (\ref{ffball result}) for the fermionic basketball.

% ==========================================================================

\section{The pieces of $I^\f_\sun$}
\label {sunset appendix}

As mentioned earlier, the fermionic setting-sun integral $I^\f_\sun$
vanishes when particle masses are ignored.  However, it is useful to
also know the piece corresponding to
\begin {equation}
   \sumint {\delta_{p_0^{}} \over P^2} \Pif(P)
   = - \sumint {1-\delta_{p_0^{}} \over P^2} \Pif(P) \,.
\end {equation}
This can be easily evaluated by relating it to the comparable bosonic piece
using (\ref{Pif Pib relation}) and then scaling momenta by 2:
\begin {eqnarray}
   \sumint {\delta_{p_0^{}} \over P^2} \Pif(P)
   &=& \sumint {\delta_{p_0^{}} \over P^2} \left[
         2^{5-d} \Pib(2P) - \Pib(P) \right]
\nonumber \\
   &=& \left(2^{2(4-d)} - 1\right) \sumint {\delta_{p_0^{}} \over P^2} \Pib(P)
\nonumber \\
   &=& \left(2^{4\eps}-1\right) \sumint_{PQ}
          {\delta_{p_0^{}} (1-\delta_{q_0^{}}) \over P^2 Q^2 (P+Q)^2}
\nonumber \\
   &=& - {T^2\over (4\pi)^2} \ln 2 + O(\eps) \,,
\end {eqnarray}
where the last equality follows from the bosonic result of
ref.~\cite{Arnold&Zhai} that
\begin {equation}
   \sumint {\delta_{p_0^{}} (1-\delta_{q_0^{}}) \over
           P^2 Q^2 (P+Q)^2}
    = {T^2 \over (4 \pi)^2}
	\left [-{1 \over 4 \epsilon}
	+ \ln {2T \over \bar\mu}
	- {1 \over 2} \right ] + O (m, \epsilon) \,.
\label {H344}
\end {equation}

% ==========================================================================

\section {Derivations of $I_\qcd^\xff$ and $I_\qcd^\xbf$}
\label {qcd appendix}

The calculations of $I_\qcd^\xff$ and $I_\qcd^\xbf$ also closely parallel
those of the purely bosonic case in Appendix H
of ref.~\cite{Arnold&Zhai}.
In that previous work, we found it algebraically convenient to rewrite the
bosonic contribution $\Pib_{\mu\nu}$ of fig.~\ref{fig pi}(a)
to the vector self-energy in terms
of
\begin {equation}
   \bar\Pib_{\mu\nu} \equiv
   2 \delta_{\mu\nu} \sumint_Q {1\over Q^2}
   - \sumint_Q {(2Q+P)_\mu (2Q+P)_\nu \over Q^2 (P+Q)^2} \,.
\label {Pib qed defn}
\end {equation}
via
\begin {equation}
   \left(\Pib_{\mu\nu}\right)^{ab}(P) =
   \ca g^2 \delta^{ab} \left[
       {d-2\over2} \bar\Pib_{\mu\nu}(P)
       - 2 (P^2\delta_{\mu\nu}-P_\mu P_\nu) \sumint_Q{1\over Q^2(P+Q)^2}
   \right] \,.
\label {Piqcdb decomposition}
\end {equation}
In order to make our presentation of the fermionic case as similar as
possible to the bosonic one, we shall do the same for the fermionic
contribution $\Pif_{\mu\nu}$ of fig.~\ref{fig pi}(b):%
\footnote{
   The main advantage to this is simplicity of presentation; one could
   just as easily do the calculation with $\Pif_{\mu\nu}$ directly.
}
\begin {equation}
   \left(\Pif_{\mu\nu}\right)^{ab}(P) =
   - \sf g^2 \delta^{ab} \left[
       2 \bar\Pif_{\mu\nu}(P)
       - 2 (P^2\delta_{\mu\nu}-P_\mu P_\nu)
	\sumint_{\{Q\}} {1\over Q^2(P+Q)^2}
   \right] \,,
\label {Piqcdf decomposition}
\end {equation}
where
\begin {equation}
   \bar\Pif_{\mu\nu} \equiv
   2 \delta_{\mu\nu} \sumint_{\{Q\}} {1\over Q^2}
   - \sumint_{\{Q\}} {(2Q+P)_\mu (2Q+P)_\nu \over Q^2 (P+Q)^2} \,.
\label {Pif qed defn}
\end {equation}
And now let's focus on computing
\begin {equation}
   I^\xff_\sqed \equiv \sumint_P {[\Delta\bar\Pif_{\mu\nu}(P)]^2 \over P^4}
   \,, \qquad
   I^\xbf_\sqed \equiv \sumint_P {\Delta\bar\Pib_{\mu\nu}(P)
                                  \Delta\bar\Pif_{\mu\nu}(P) \over P^4} \,.
\label {Isqedf defn}
\end {equation}
Following through the same steps as in Appendix H of
ref.~\cite{Arnold&Zhai}, one gets
the obvious generalizations of eqs.~(H27-29) of that reference:
\begin {eqnarray}
   \sumint_P {1\over P^4} \left[\Delta\bar\PifT_{\mu\nu}(P)\right]^2
   &=& \sumint_P \left[ \PifT(P) \right]^2
     + 4(d-2) b_2 f_1^2
     + O(\eps) \,,
\\
   \sumint_P {1\over P^4} \Delta\bar\PibT_{\mu\nu}(P)
	\Delta\bar\PifT_{\mu\nu}(P)
   &=& \sumint_P \PibT(P) \PifT(P)
     + 4(d-2) b_2 b_1 f_1
     + O(\eps) \,,
\end {eqnarray}
\begin {eqnarray}
   \sumint_P {1\over P^4} \bar\PifT_{\mu\nu}(P) \bar\PiZ_{\mu\nu}(P)
   &=& {1\over d-1} \sumint_P \PifT(P) \PiZ(P)
      + 2 {(d-2)\over(d-1)} f_1 \sumint_P {1\over P^2} \PiZ (P)
   \,,
\\
   \sumint_P {1\over P^4} \bar\PibT_{\mu\nu}(P) \bar\PiZ_{\mu\nu}(P)
   &=& {1\over d-1} \sumint_P \PibT (P)\PiZ(P)
      + 2 {(d-2)\over(d-1)} b_1 \sumint_P {1\over P^2} \PiZ(P)
   \,,
\label {sqed T0 result}
\end {eqnarray}
\begin {equation}
   \sumint_P {1\over P^4} \left[ \bar\PiZ_{\mu\nu}(P)\right]^2
   = {1\over d-1} \sumint_P \left[ \PiZ (P)\right]^2 \,.
\label {sqed 00 result}
\end {equation}
Summing these results, and incorporating the results for the assorted
basic integrals, gives
\begin {eqnarray}
   I_\sqed^\xff
   &=& {1\over(4\pi)^2} \left(T^2\over12\right)^2 \left[
        {11\over6\eps} + 11 \lnmub
        + {1\over3} {\zeta'(-3)\over\zeta(-3)}
        + {20\over3} {\zeta'(-1)\over\zeta(-1)}
        + 4 \gammaE
        + {281\over60}
        - 13 \ln 2   \right]
\nonumber \\ && \qquad\qquad\qquad\qquad
   + O(\eps) \,,
\\
   I_\sqed^\xbf
   &=& {1\over(4\pi)^2} \left(T^2\over12\right)^2 \left[
        -{59\over12\eps} - {59\over2} \lnmub
        + {11\over6} {\zeta'(-3)\over\zeta(-3)}
        - {70\over3} {\zeta'(-1)\over\zeta(-1)}
        - 8 \gammaE
        - {2063\over120}
        + {169\over10} \ln 2   \right]
\nonumber \\ && \qquad\qquad\qquad\qquad
   + O(\eps) \,.
\label {Isqed result}
\end {eqnarray}
Using (\ref{Piqcdb decomposition}) and (\ref{Piqcdf decomposition}) and our
standard reduction tricks yields:%
\footnote{
  Because of our slightly different methods of bookkeeping
  of infrared divergences,
  our reduction of $I_\qcd^\xff$ differs from a similar
  reduction in ref.~\cite{Coriano&Parwani} by
  $M_4^2 \textsumint (1-\delta_{p_0^{}}) P^{-2} \Pif(P)$.
  There is a canceling difference in our treatment of diagram (o) of
  fig.~\ref{figf2}.
}
\begin {eqnarray}
   I_\qcd^\xff &=&
      4 I_\sqed^\xff
      + 4(d-3) I_\bball^\xff
      - 16(d-2) f_1 \sumint' {1\over P^2} \Pif(P)
\nonumber \\
   &=& {1\over(4\pi)^2} \left(T^2\over12\right)^2 \left[
        {40\over3\eps} + 80 \lnmub
        - {32\over3} {\zeta'(-3)\over\zeta(-3)}
        + {224\over3} {\zeta'(-1)\over\zeta(-1)}
        + 16 \gammaE
        + {124\over3}
        + {448\over5} \ln 2   \right]
\nonumber \\ && \qquad\qquad\qquad\qquad
   + O(\eps) \,,
\\
   I_\qcd^\xbf &=&
      (2-d) I_\sqed^\xbf
      - 3(d-2) I_\bball^\xbf
      + 8(d-2) f_1 \sumint_P' {1\over P^2} \Pib(P)
      + 2(d-2)^2 b_1 \sumint_P' {1\over P^2} \Pif(P)
\nonumber \\
   &=& {1\over(4\pi)^2} \left(T^2\over12\right)^2 \Biggl[
        - {29\over3\eps} - 58 \lnmub
        - {38\over3} {\zeta'(-3)\over\zeta(-3)}
        + {104\over3} {\zeta'(-1)\over\zeta(-1)}
\nonumber \\ && \qquad\qquad\qquad\qquad
        + 16 \gammaE
        - {251\over10}
        + {398\over5} \ln 2
        - 96 \ln(2\pi)   \Biggr]
   + O(\eps) \,.
\end {eqnarray}

Before leaving this section, we note that one can use our results to
derive the basic integral
\begin {equation}
   H_4 \equiv \sumint_{P\{QK\}}
     {(Q\cdot K)^2 \over P^4 Q^2 K^2 (P+Q)^2 (P+K)^2}
\end {equation}
defined by Parwani and Corian\`o.  By applying our usual reduction techniques
to $I_\sqed^\xff$, and using the fact that $I_\sun^\f = 0$,
one can derive that
\begin {equation}
   I_\sqed^\xff = 16 H_4 - I_\bball^\xff + 4(d-4) b_2 f_1^2 \,.
\end {equation}
Solving for $H_4$ and using our results for $I_\sqed^\xff$ and
$I_\bball^\xff$ then yields
\begin {equation}
   H_4 = {1\over(4\pi)^2} \left(T^2\over12\right)^2 \left[
        {5\over24\eps} + {5\over4} \lnmub
        - {1\over6} {\zeta'(-3)\over\zeta(-3)}
        + {7\over6} {\zeta'(-1)\over\zeta(-1)}
        + {1\over4} \gammaE
        + {23\over24}
        - {8\over5} \ln 2   \right]
   + O(\eps) \,,
\end {equation}
which agrees with the numerical result of Parwani and Corian\`o within errors.


\begin{references}

\bibitem{Arnold&Zhai}
    P. Arnold and C. Zhai,
        University of Washington preprint UW/PT-94-03, hep-ph/9408276.

\bibitem{Coriano&Parwani}
    C. Corian\`o and R. Parwani,
        Argonne National Lab preprint ANL-HEP-PR-94-02 (revised) (1994),
	  hep-ph/9405343;
    R. Parwani, Saclay preprint SPhT/94-065 (1994), hep-ph/9406318;
    R. Parwani and C. Corian\`o,
        Argonne National Lab preprint ANL-HEP-PR-94-32 (1994), hep-ph/9409269.

\end {references}
\end {document}